\documentclass[twocolumn,floatfix,showpacs,superscriptaddress]{revtex4}
\usepackage{graphicx}% Include figure files
\usepackage{bm}% bold math

\begin{document}

\title{Loss and revival of phase coherence in a Bose-Einstein condensate\\
moving through an optical lattice}

\author{Francesco Nesi}
\email{nesi@lens.unifi.it}
\affiliation{%
  LENS - Dipartimento di Fisica, Universit\`a di Firenze and INFM\\
  Via Nello Carrara 1, 50019 Sesto Fiorentino, Italy }
\author{Michele Modugno}
\email{modugno@fi.infn.it}
\affiliation{%
  LENS - Dipartimento di Fisica, Universit\`a di Firenze and INFM\\
  Via Nello Carrara 1, 50019 Sesto Fiorentino, Italy }
\affiliation{%
  BEC-INFM Center, Universit\`a di Trento, 38050 Povo, Italy}

\date{\today}

\begin{abstract}
We investigate the phase coherence of a trapped Bose-Einstein
condensate that undergoes a dynamical superfluid-insulator transition
in the presence of a one-dimensional optical lattice. We study the
evolution of the condensate after a sudden displacement of the
harmonic trapping potential by solving the Gross-Pitaevskii equation,
and comparing the results with the prediction of two effective 1D
models. We show that, owing to the 3D nature of the system, the
breakdown of the superfluid current above a critical displacement is
not associated to a sharp transition, but there exists a range of
displacements for which the condensate can recover a certain degree of
coherence. We also discuss the implications on the interference
pattern after the ballistic expansion as measured in recent 
experiments at LENS.
\end{abstract}
\maketitle

%===============================================================
\section{Introduction}
%===============================================================

The possibility of manipulating Bose-Einstein condensates (BECs) in
periodic potentials has provided the opportunity to investigate a wide
range of phenomena, exploring a very stimulating field which combines
atomic and solid state physics
\cite{anderson,orzel,pedri,cataliotti,greiner,burger,lens,morsch}.

In particular, the dynamical behavior and coherence properties of BECs
loaded in optical lattices have been the subject of an extensive
experimental \cite{pedri,cataliotti,burger,lens,morsch} and theoretical work 
\cite{smerzi,wuniu,pethick,salerno}.  
Recently it has been demonstrated
that a trapped BEC moving through a one-dimensional optical lattice
created by a laser standing wave can realize a Josephson junction
array sustaining an oscillating atomic current \cite{cataliotti}. The
condensate is set in motion by a sudden displacement of the trapping
potential, and then, when the center-of-mass velocity reaches a
critical value, the system undergoes a transition from the superfluid
regime to an insulator regime, characterized by a localization of the
condensate in the trapping potential \cite{lens}.

This phenomenon, which is accompanied by a loss of coherence, is
triggered by the onset of a dynamical instability, as theoretically
demonstrated in \cite{smerzi,wuniu,pethick}. These studies rely on the
analysis of the Bogoliubov spectrum of simplified one-dimensional
models, and predict a sharp transition from the superfluid to the
insulator regime, with a complete loss of the system coherence.  Such
a behavior has been confirmed in the deep insulator regime, that is
for large initial displacements, by the direct solution of the three
dimensional Gross-Pitaevskii equation (GPE-3D) \cite{adhikari}.

In this work we show that, owing to the 3D character of the system, a
non-trivial phenomenology takes place in the intermediate regime where
the condensate is able to retain a certain degree of coherence. The
role played by the dimensionality of the system is investigated by
comparing explicitly the solution of the GPE-3D with those of two
one-dimensional models. We also discuss how the phase coherence of the
system influences the interference pattern after the ballistic
expansion, showing that the interference peaks which characterize the
superfluid phase \cite{pedri} are not completely destroyed in the
insulator regime, though they may have a reduced visibility. The onset of
decoherence processes is signaled also by additional structures
(``fringes'') which appear in the central peak, as recently observed
at LENS \cite{lens,fort}.

The paper is organized as follows. In the next section we introduce
the GPE-3D and the two one-dimensional models considered
(GPE-1D and NPSE). Then, in section \ref{sec:results} we present the
results by discussing the evolution of the phase coherence, the
Fourier power spectrum of the system, the center-of-mass dynamics, and
eventually the effect of the free expansion of the system. The
conclusions are drawn in section \ref{sec:conclusions}.

%===============================================================
\section{General Formulation}
\label{sec:general}
%===============================================================

Let us consider a Bose-Einstein condensate trapped in a cylindrically
symmetric harmonic potential superimposed to an optical lattice.  The
dynamics of the system is described by the three-dimensional 
Gross-Pitaevskii equation (GPE-3D) \cite{review}
\begin{equation} 
\label{eq:gpe-3d}
i\hbar \frac{\partial}{\partial t}\Psi(\mathbf{x},t)=
\biggl[-\frac{\hbar^2}{2m}\nabla^2+V(\mathbf{x})+
     gN|\Psi|^2\biggr]\Psi(\mathbf{x},t)
\end{equation}
where $N$ is the number of condensed atoms, $g=4\pi\hbar^2a/m$ the
coupling strength, $m$ the atomic mass and $a$ the inter-atomic
scattering length.  The external potential
$V(\textbf{x})=V_{ho}(\textbf{x})+V_p(z)$ is the sum of the harmonic
trapping potential
\begin{equation}
V_{ho}(\mathbf{x})=\frac{1}{2} m \left( \omega_{\perp}^2
r^2+\omega_z^2 z^2 \right)
\end{equation}
and of the periodic potential generated by the optical lattice
along the axial direction $z$
\begin{equation}
V_{p}(z)=s\, E_r \cos^2\left(\frac{2\pi z}{\lambda}\right)\,,
\end{equation}
$\lambda$ being the wavelength of the laser, $E_r\equiv
h^2/2m\lambda^2$ the recoil energy of an atom absorbing
one lattice photon, $d=\lambda/2$ the distance between two adjacent
minima (lattice sites), and $s$ a dimensionless parameter controlling
the intensity of the lattice.

To model the LENS experiment \cite{lens}, we consider a
$^{87}\mathrm{Rb}$ condensate with $N=5\cdot10^4 $ atoms, and the
following parameters characterizing the external potential:
$\omega_\perp = 2\pi\cdot92$ Hz, $\omega_z = 2\pi\cdot9$ Hz, $
\lambda=795 $\,nm, and $s=5$.

\begin{figure}
\centerline{\includegraphics[width=8cm,clip=]{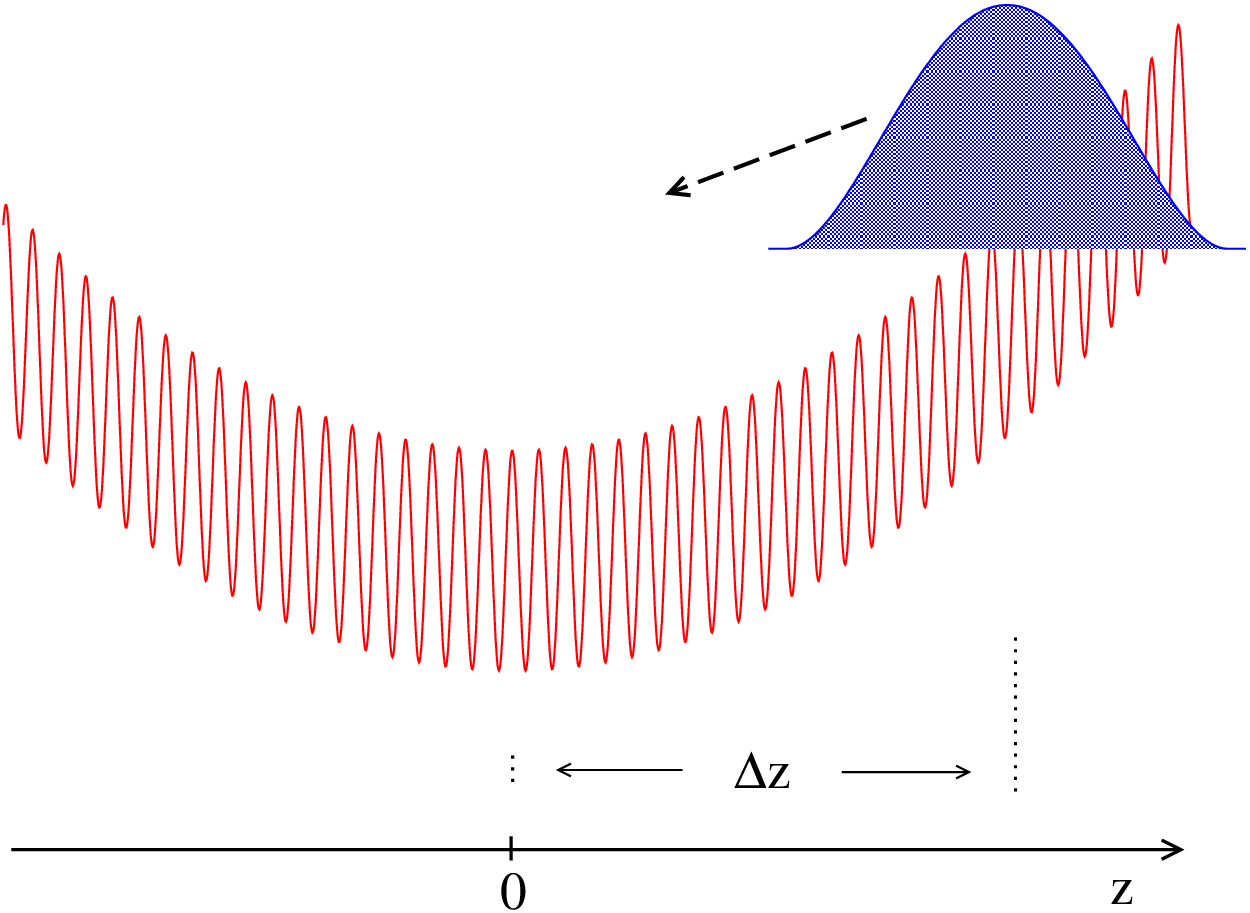}}
\caption{Schematic of the system setup.}
\label{fig:scheme}
\end{figure}

The ground state of the system in the combined harmonic plus periodic
potential is found by mapping the wave function on a discretized
grid \cite{sites} and using a standard imaginary time evolution
\cite{review}.  Then, at $t=0$, the harmonic potential is shifted by
$\Delta z$ and the condensate is set to evolve in the combined potentials
as sketched in Fig. \ref{fig:scheme}. To solve the time-dependent
GPE-3D we use a split-step method which combines a FFT evolution
in the axial direction \cite{splitstep,numrec} and a Crank-Nicholson
algorithm for the radial one \cite{brunello}.

In order to investigate the role played by the transverse dimension we
also compare the full three-dimensional behavior of the GPE-3D with
the solutions of two one-dimensional effective models that account for
the axial dynamics along the lattice direction.

The first model considered is the 1D Gross-Pitaevskii equation
(GPE-1D) given by
\begin{equation}
i\hbar \frac{\partial}{\partial t}\psi(z,t)=
\left[-\frac{\hbar^2}{2m}\nabla_z^2+V(z)+
g_{1D} N|\psi|^2\right]\psi(z,t)
\end{equation}
with
\begin{equation}
V(z)=\frac{1}{2} m \omega_z^2 z^2 + V_{p}(z)
\end{equation}
where  $g_{1D}$ is obtained by a suitable renormalization of the 3D
coupling constant $g$. In particular, by requiring the invariance of
the Thomas-Fermi (TF) chemical potential (that is the invariance of 
axial size of the condensate in the TF limit) \cite{band}, we get
\begin{equation}
g_1=\frac{4}{3N}\sqrt{2}\,\hbar\omega_z{\left(\frac{\mu_{TF}}
{\hbar\omega_z}\right)}^{3/2}\!a_z 
\label{eq:g1}
\end{equation}
with
\begin{equation}
\mu_{TF}=\frac{1}{2}\hbar\omega_\perp\left(15
\frac{\omega_z}{\omega_\perp}N
\frac{a_s}{a_\perp}\right)^{2/5}
\end{equation}
where $a_z=\sqrt{\hbar/(m\,\omega_z)}$ and
$a_\perp=\sqrt{\hbar/(m\,\omega_\perp)}$ are the characteristic oscillator 
lengths.

The other model considered here is described by the Non-Polynomial
Schr\"odinger equation (NPSE) \cite{salasnich} obtained from the
GPE-3D by means of a factorization of the condensate wavefunction
in the product of a gaussian radial component of width $\sigma(z,t)$,
and of an axial wavefunction $\psi(z,t)$ that satisfies the 
differential equation (NPSE) 
\begin{eqnarray} 
i\hbar\frac{\partial}{\partial
t}\psi(z,t)&=&
\bigg[-\frac{\hbar^2}{2m}\nabla_z^2+V(z)+\frac{gN}{2\pi\sigma^2}
|\psi|^2\\
&&+\frac{1}{2}\hbar \omega_\perp\left(\frac{a_\perp^2}{\sigma^2}+
\frac{\sigma^2}{a_\perp^2}\right)\bigg]\psi(z,t)\,,
\nonumber
\end{eqnarray}
coupled with an algebraic equation for the radial width 
\begin{equation}
\sigma(z,t)=a_\perp\sqrt[4]{1+2a_sN|\psi(z,t)|^2}\,.
\end{equation}
Owing to this partial coupling between axial and radial degrees of
freedom, the NPSE is expected, with respect to the GPE-1D, to provide
a more accurate description of the actual ground-state and dynamics of
the system at least in the coherent regime, as discussed in
\cite{salasnich,massignan}.

%===============================================================
\section{Results and discussion}
\label{sec:results}
%===============================================================

In this section we investigate the phase coherence and the
center-of-mass dynamics of the condensate during the trapped evolution
through the optical lattice.  To discuss the degree of coherence of
the system we also show the evolution of the momentum distribution of
the condensate and the expected signatures after the free expansion of
the system.

%-------------------------------------------------
\subsection{Coherence}
%-------------------------------------------------

In order to characterize the degree of coherence of the condensate
inside the optical lattice we define the quantity $\chi(t)$ 
as the squared modulus of the correlation between values of the condensate
wave function evaluated at points corresponding to the distance between
nearest-neighbor sites of the lattice \cite{smerzi}
\begin{equation} 
\chi(t)\equiv\left|\int\! d^3x\,
\Psi^\ast(r,z;t)\,\Psi(r,z+d;t)\right|^2\,.
\label{eq:chidef}
\end{equation}
In the following we will refer to $\chi$ as ``coherence'':
$\chi=1$ means that the system is fully coherent, whereas $\chi=0$ 
indicates a complete decoherence and the loss of the superfluid properties.

\begin{figure} 
\centerline{\includegraphics[width=8cm,clip=]{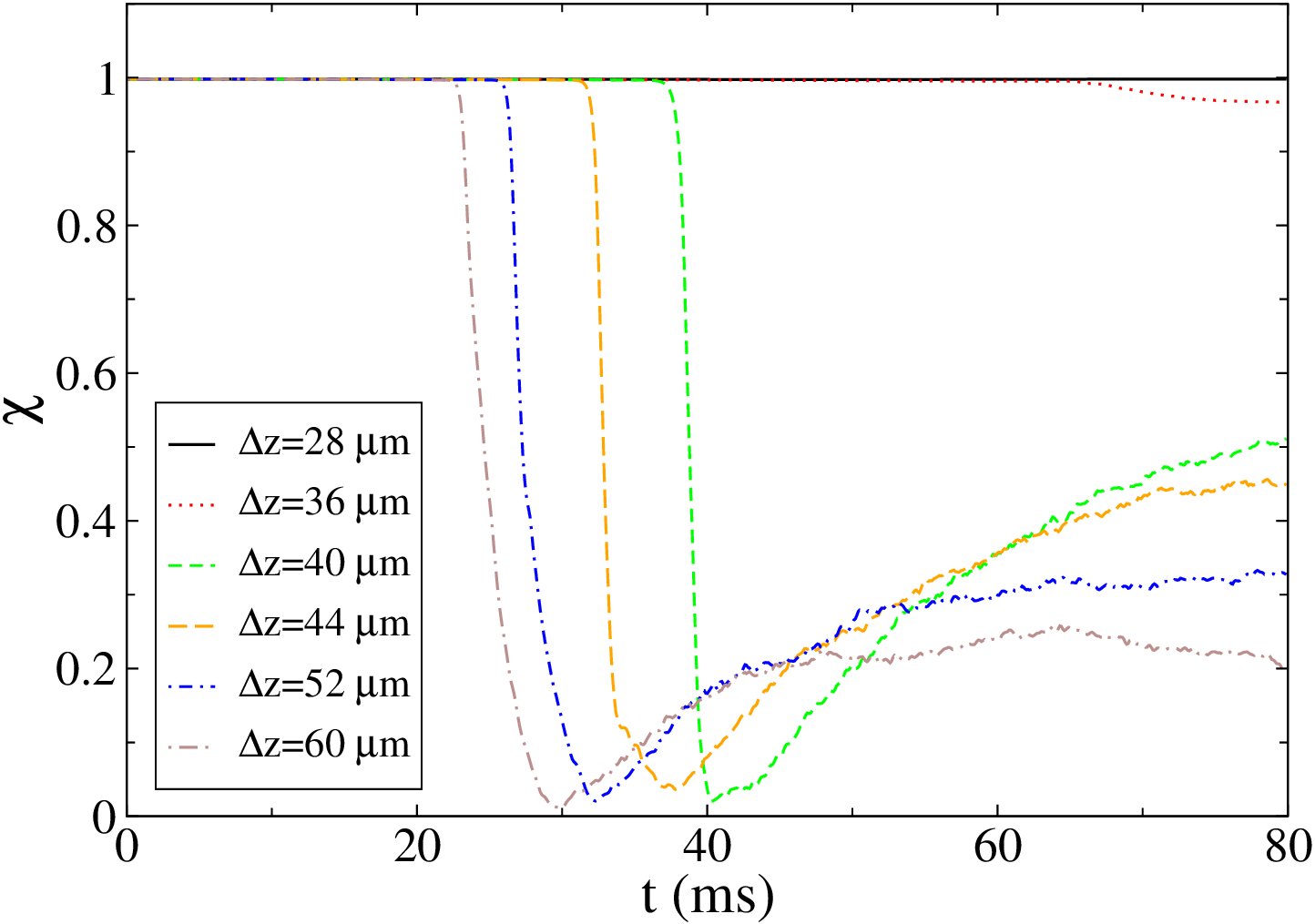}}
\caption{Coherence of the condensate between nearest-neighbor sites of
the optical lattice, obtained from the solution of the GPE-3D, for
several initial displacements.  After the onset of the
dynamical instability (for $\Delta z\simeq40\,\mu$m) the system
undergoes a sudden loss of coherence, that is partially recovered for
later times.}
\label{fig:coherence-3d}
\end{figure}

In Fig. \ref{fig:coherence-3d} we show the evolution of the coherence
for several initial displacement, up to $80$\,ms.
The picture shows that above a critical displacement, after an initial
coherent evolution, the system undergoes a sudden loss of coherence,
that is partially recovered in the subsequent evolution. We have
verified that for longer times (up to $200$\, ms) the value of $\chi(t)$
does not grow anymore and remains close to the ``saturation value'' at
$80$\,ms.

We note that even though the system may become completely incoherent
($\chi=0$), coherence is preserved in each sub-condensate (in each
lattice site), and this is sufficient to justify our description of
the system in terms of solutions of a differential equation.

\begin{figure} 
\centerline{\includegraphics[width=8cm,clip=]{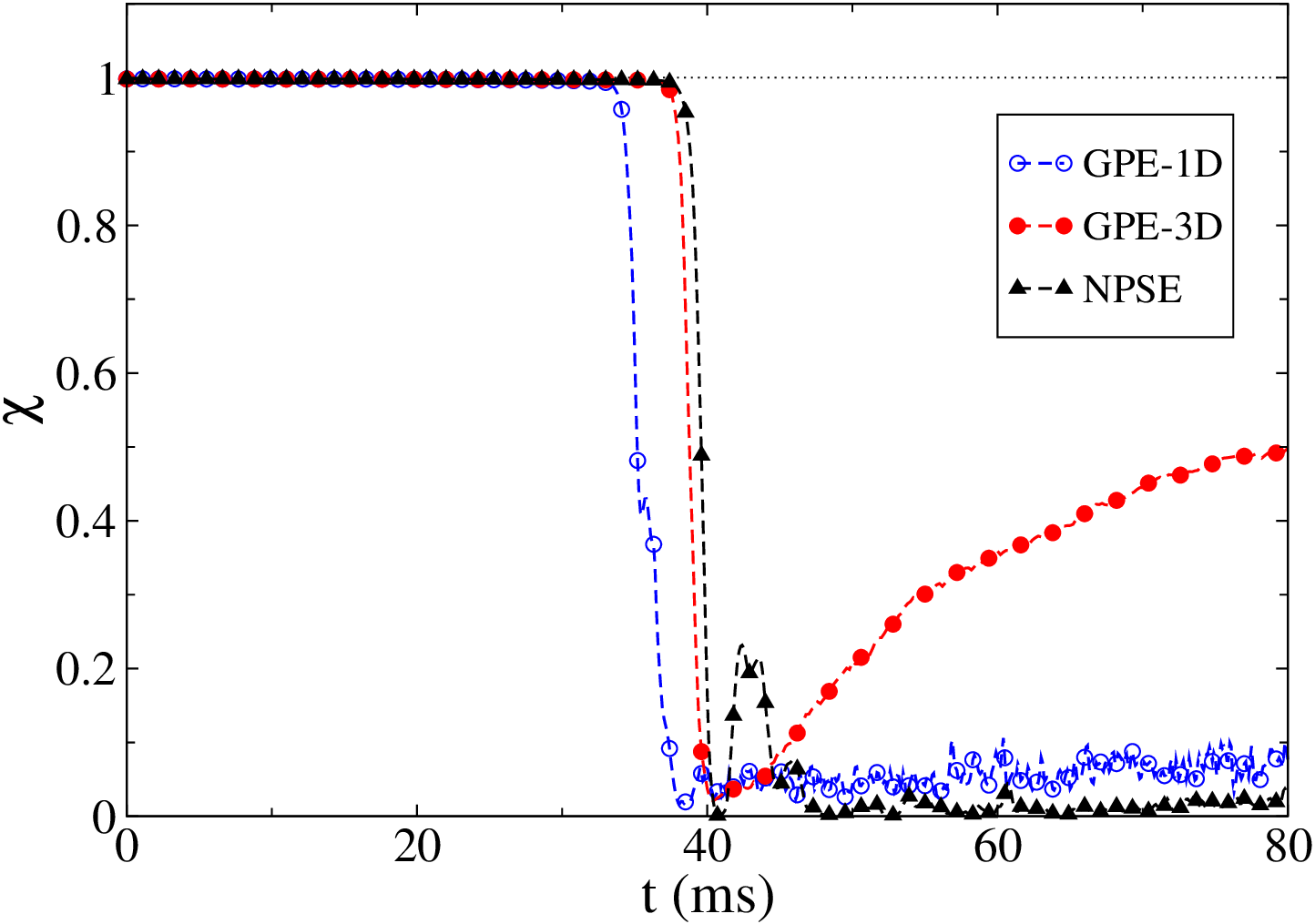}}
\caption{Coherence of the condensate between nearest-neighbor sites of
the optical lattice for a displacement $\Delta z=40\,\mu$m.  
When the system enter the insulator regime the
1D models (GPE-1D, empty circles; NPSE, triangles) predict a complete
loss of coherence, while the actual (3D) behavior of the system is
characterized by a partial recover of coherence
(GPE-3D, filled circles).}
\label{fig:coherence}
\end{figure}

As shown in Fig. \ref{fig:coherence} this capability of the system to
regain a certain degree of coherence is due to its 3D nature. In
fact, both the NPSE and the GPE-1D predict a complete dephasing after
the onset of the instability, characterized by an almost vanishing coherence.
The failure of these 1D effective models in the insulator regime is
due to the fact that at the onset of the instability most of the
energy initially associated to the coherent evolution of the system is
absorbed by the modes responsible for the dynamical instability, which grow
exponentially in time destroying the phase coherence of the condensate 
\cite{wuniu,smerzi}. However, while in the 1D case all the energy is
transferred to these modes, in the 3D case the interplay between
radial and axial degrees of freedom, coupled through the nonlinear
term, allows for a partial transfer of energy to coherent modes of the
condensate, acting, as a matter of fact, as a restraint to the
decoherence process
\footnote{The investigation of actual mechanism which determines
this behavior deserves a deeper investigation that goes beyond the 
scope of this work.}.
Note in this respect that the partial coupling between axial and
radial degrees of freedom in the NPSE is not sufficient to account for
the rephasing mechanism, and this is likely due to the simple gaussian
factorization of the wave-function which produces a coupling just
between radial and axial densities \cite{salasnich}, losing any
information on the phase dynamics \footnote{In the case of coherent
excitations, and also for a free expansion, it is possible to use a
more reliable ansatz as discussed in Ref. \cite{massignan}}.

%-------------------------------------------------
\subsection{Center-of-mass dynamics}
%-------------------------------------------------

Let us now discuss the center-of-mass evolution along the lattice direction,
by considering the axial coordinate
\begin{equation}
z_{cm}(t)=\int\! d^3x\:z\,|\Psi(r,z;t)|^2\,,
\end{equation}
and its velocity
\begin{equation}
v_{cm}(t)=\frac{\hbar}{2i\,m}\int\! 
d^3x\left[\Psi^*\nabla_z\Psi-\Psi\nabla_z\Psi^*\right]\,.
\end{equation}

\begin{figure}
\centerline{\includegraphics[width=8cm,clip=]{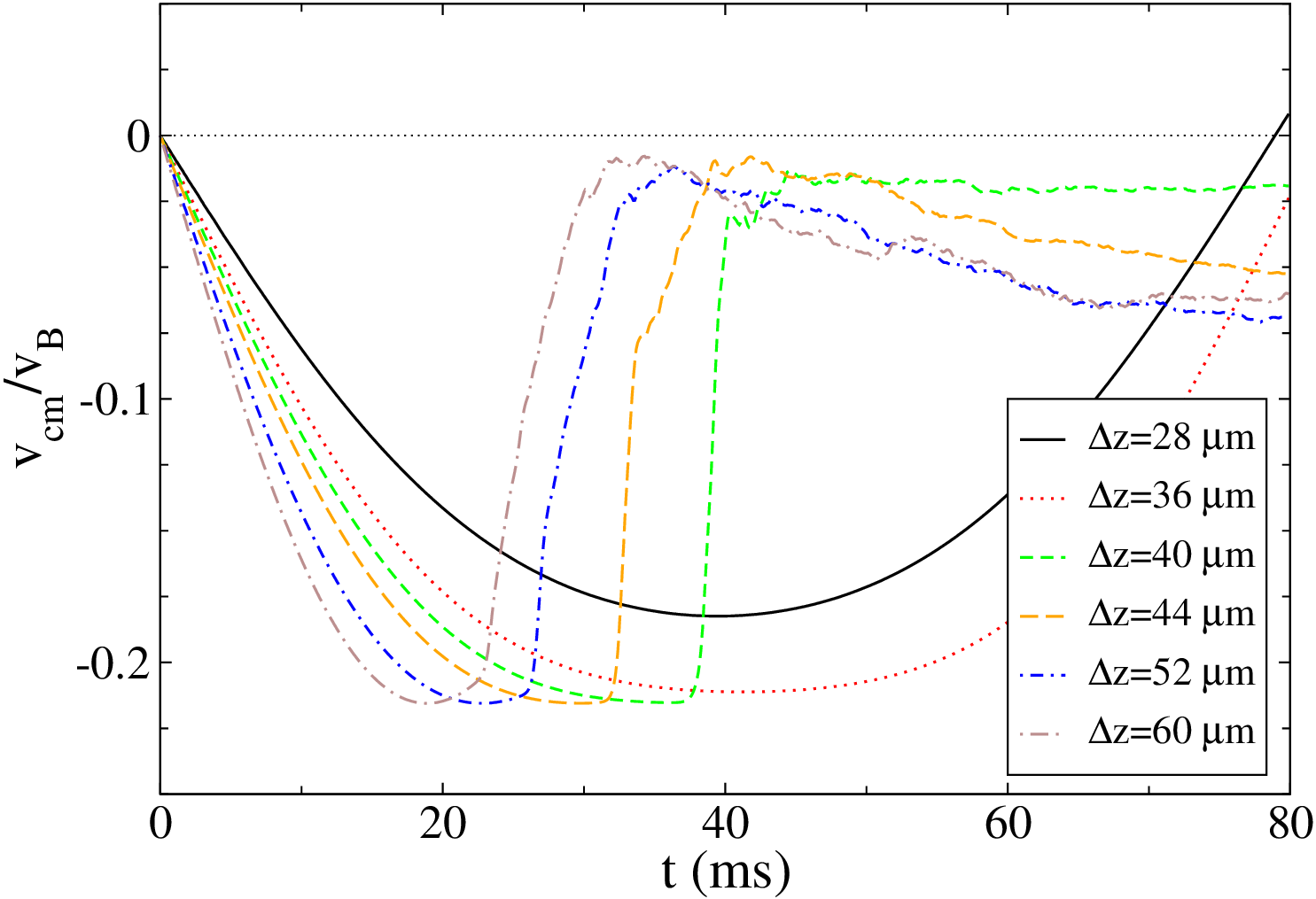}}
\caption{Center-of-mass velocity of the condensate (in units of the
Bragg velocity $v_B=\hbar\pi/(md)$) obtained from the solution of the
GPE-3D, for several initial displacements. This picture
clearly shows that the decoherence process is characterized by the
achievement of a critical velocity (see also
Fig. \ref{fig:coherence}).}
\label{fig:vm-3d}
\end{figure}

In Fig. \ref{fig:vm-3d} we show the evolution of $v_{cm}(t)$ during
the trapped dynamics, for several initial displacements
 (see also Fig. \ref{fig:coherence-3d}).  This picture points out
clearly the existence of a critical velocity independent of the initial 
displacement (here
$|v_{crit}|\simeq1.25$\,mm/s, in nice agreement with the experimental
value reported in \cite{lens}), 
beyond which the system cannot sustain a
coherent oscillation. This is due to the presence of axial modes with
imaginary frequency that grow exponentially in time and are responsible 
for the dynamical instability
of the system in a certain range of condensate quasimomenta
\cite{smerzi,wuniu}.

\begin{figure}
\centerline{\includegraphics[width=8cm,clip=]{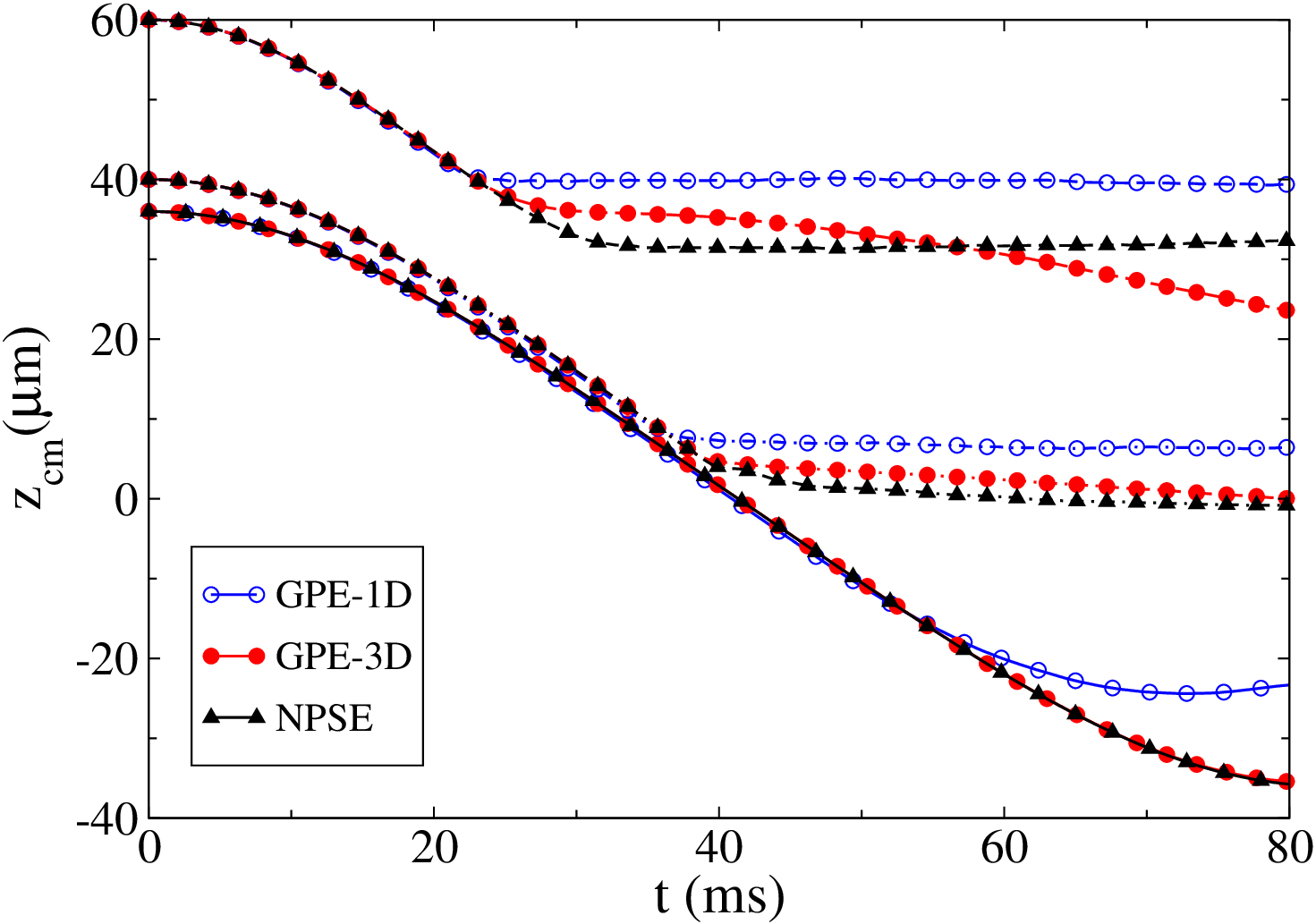}}
\caption{Center-of-mass dynamics for three values of the initial
displacement ($\Delta z=36\,\mu$m continuous line, 
$\Delta z=40\,\mu$m dashed-dotted, $\Delta z=60\,\mu$m dashed). 
Contrarily to what happens in the simplified 1D models
(GPE-1D, empty circles; NPSE, triangles), in the insulator regime
the condensate center-of-mass continues a slow evolution toward the center 
of the trap (GPE-3D, filled circles).}
\label{fig:cm}
\end{figure}

In Fig. \ref{fig:cm} we compare the corresponding center-of-mass
evolution with the prediction of the one-dimensional models, for three
initial displacements. We notice that the GPE-3D predicts a slow motion
toward the center of the trapping potential, as observed in the
experiment \cite{lens,fort}, whereas the 1D models predict a
sudden localization of the system as a consequence of the complete loss of
coherence.

Note also that for small initial displacements, that is in the regime
of coherent oscillations, the predictions of the NPSE are in nice
agreement with those obtained by the full GPE-3D, while the GPE-1D
deviates from the expected behavior after about 40 ms of evolution (we
have observed in a similar fashion that the GPE-1D slightly
underestimates the critical displacement).  This effect may be due to
the chosen renormalization of the coupling constant $g_1$ (see
Eq. (\ref{eq:g1})) that can overestimate non-linear effects
\cite{massignan} responsible for the existence of dynamically unstable
modes. However, even though one tunes $g_1$ in order to better
reproduce the actual critical displacement, this would not change the
inadequateness of the model to describe the behavior of the system in
the insulator regime, neither the fact that in any case the NPSE
represents a more reliable effective model (with respect of the
``simple'' GPE-1D) to describe the dynamics of the system in the
superfluid regime \cite{salasnich,massignan}.

%-------------------------------------------------
\subsection{Momentum evolution}
%-------------------------------------------------

\begin{figure}
\centerline{\includegraphics[width=7cm,clip=]{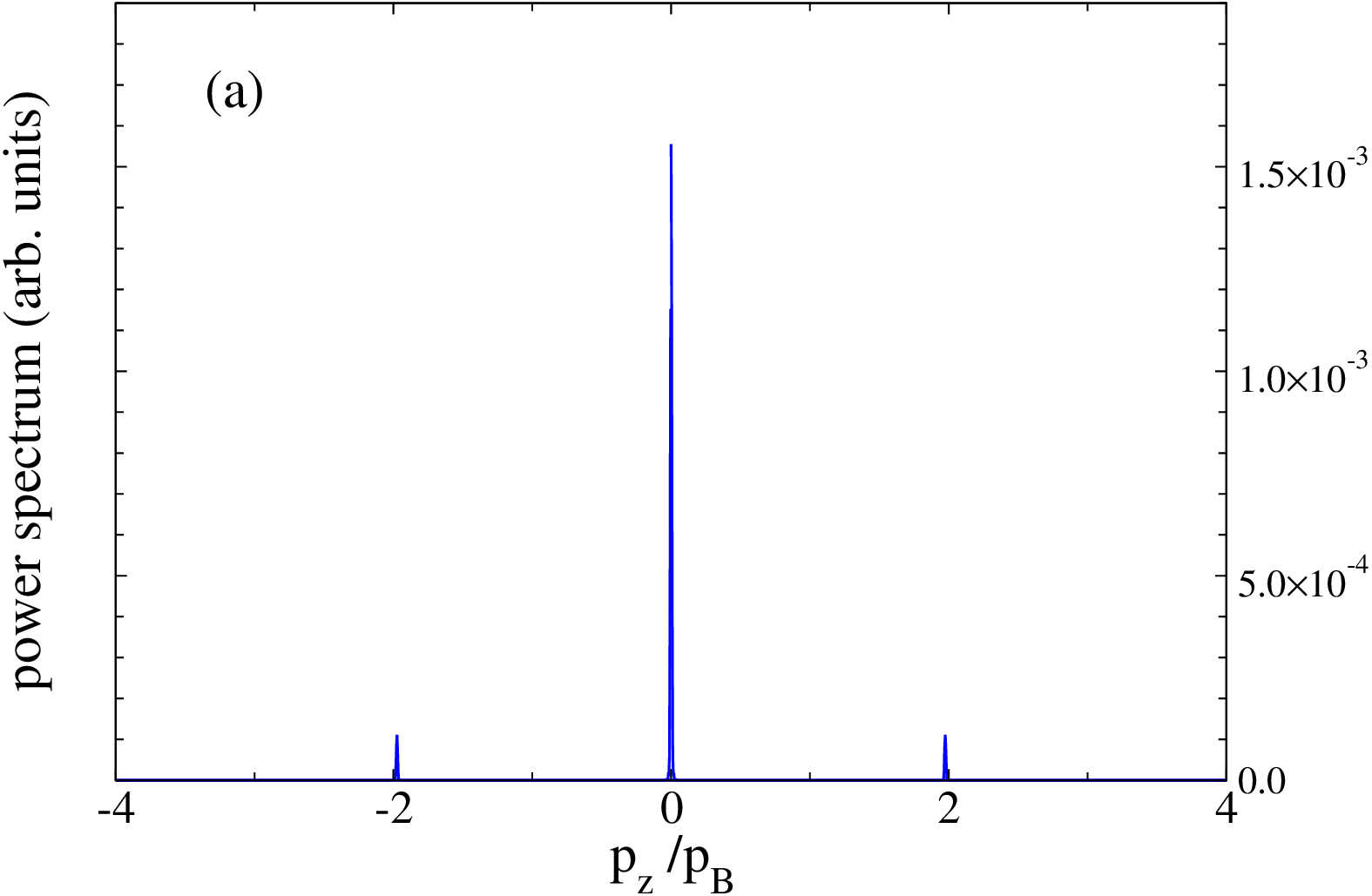}}
\centerline{\includegraphics[width=7cm,clip=]{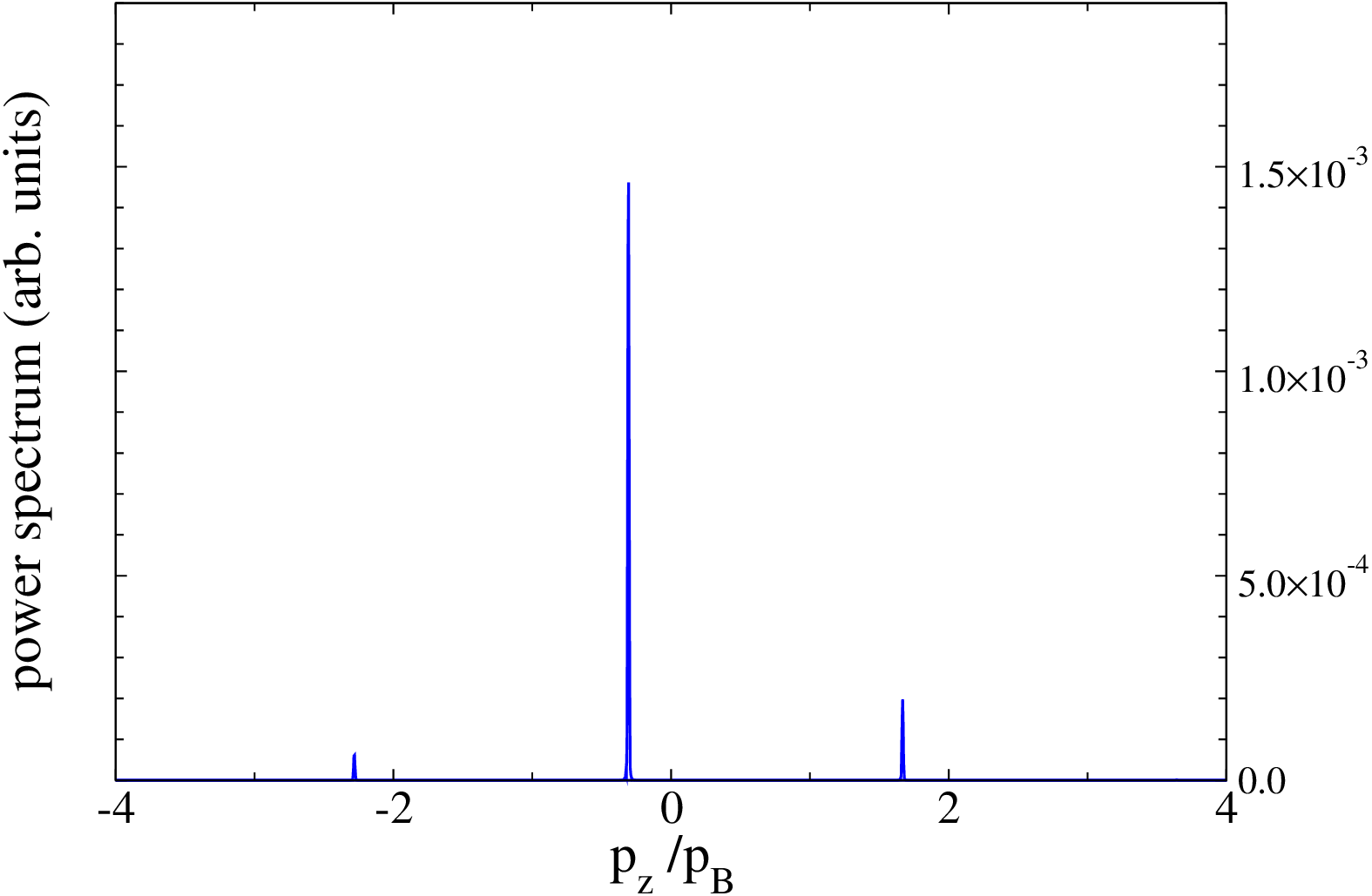}}
\centerline{\includegraphics[width=7cm,clip=]{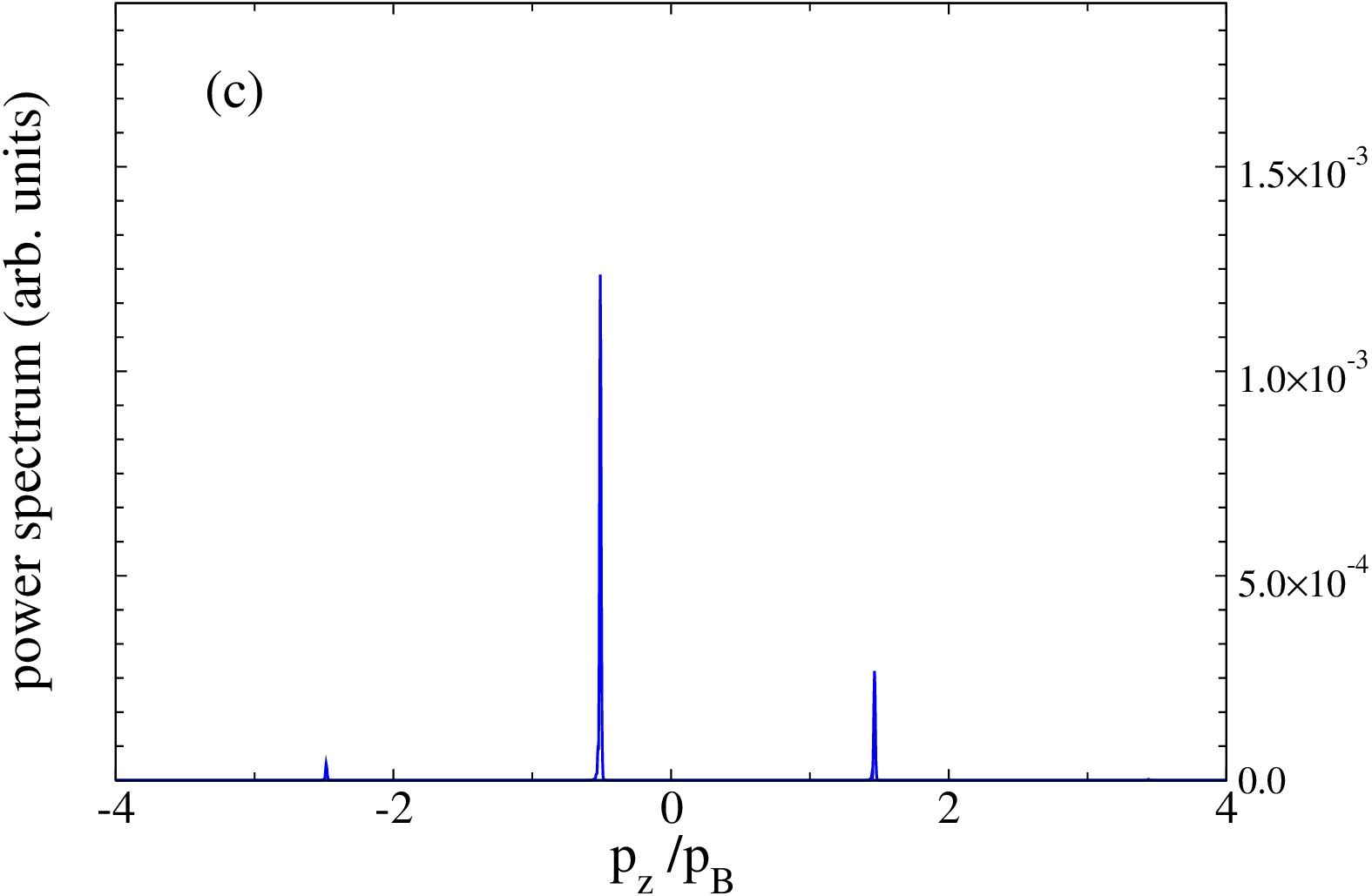}}
\caption{Axial power spectrum of the condensate in the superfluid
regime for $t=0$\,ms (a), $15$\,ms (b) and $30$\,ms (c). The presence
of sharp peaks spaced apart by multiples of $\pm2p_B$  is a
signature of the system coherence \cite{pedri}.}
\label{fig:p1}
\end{figure}

\begin{figure}
\centerline{\includegraphics[width=7cm,clip=]{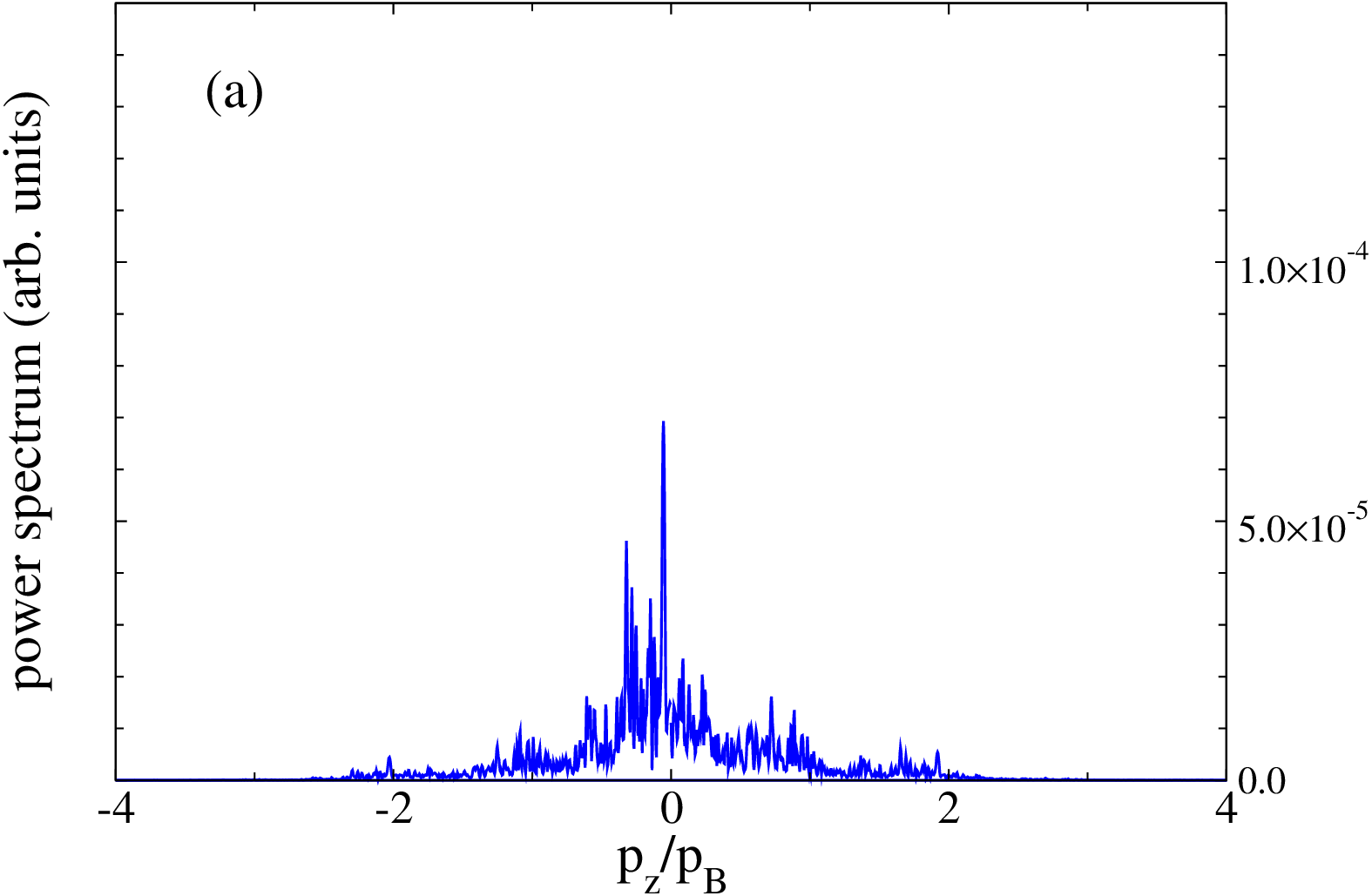}}
\centerline{\includegraphics[width=7cm,clip=]{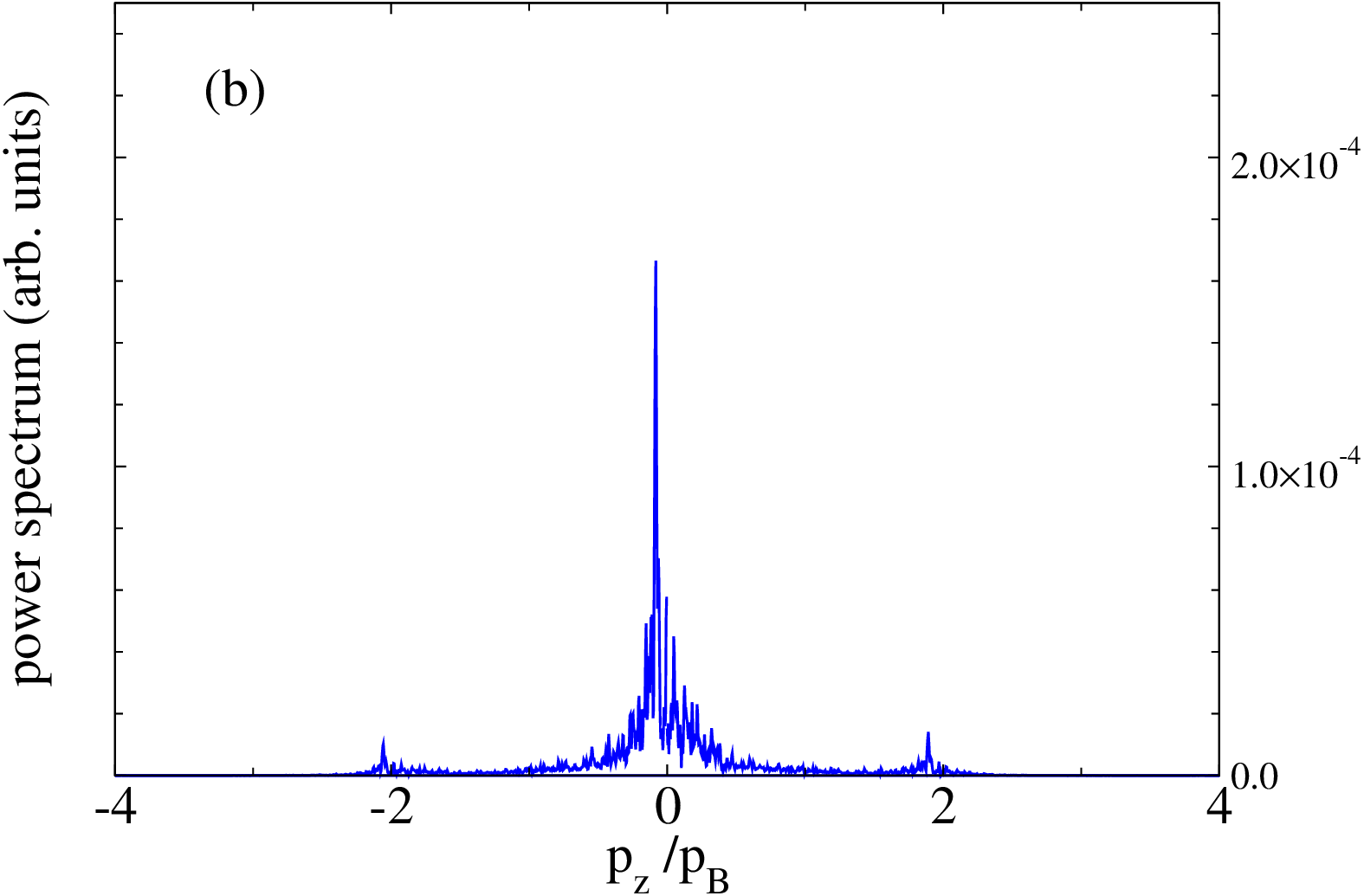}}
\centerline{\includegraphics[width=7cm,clip=]{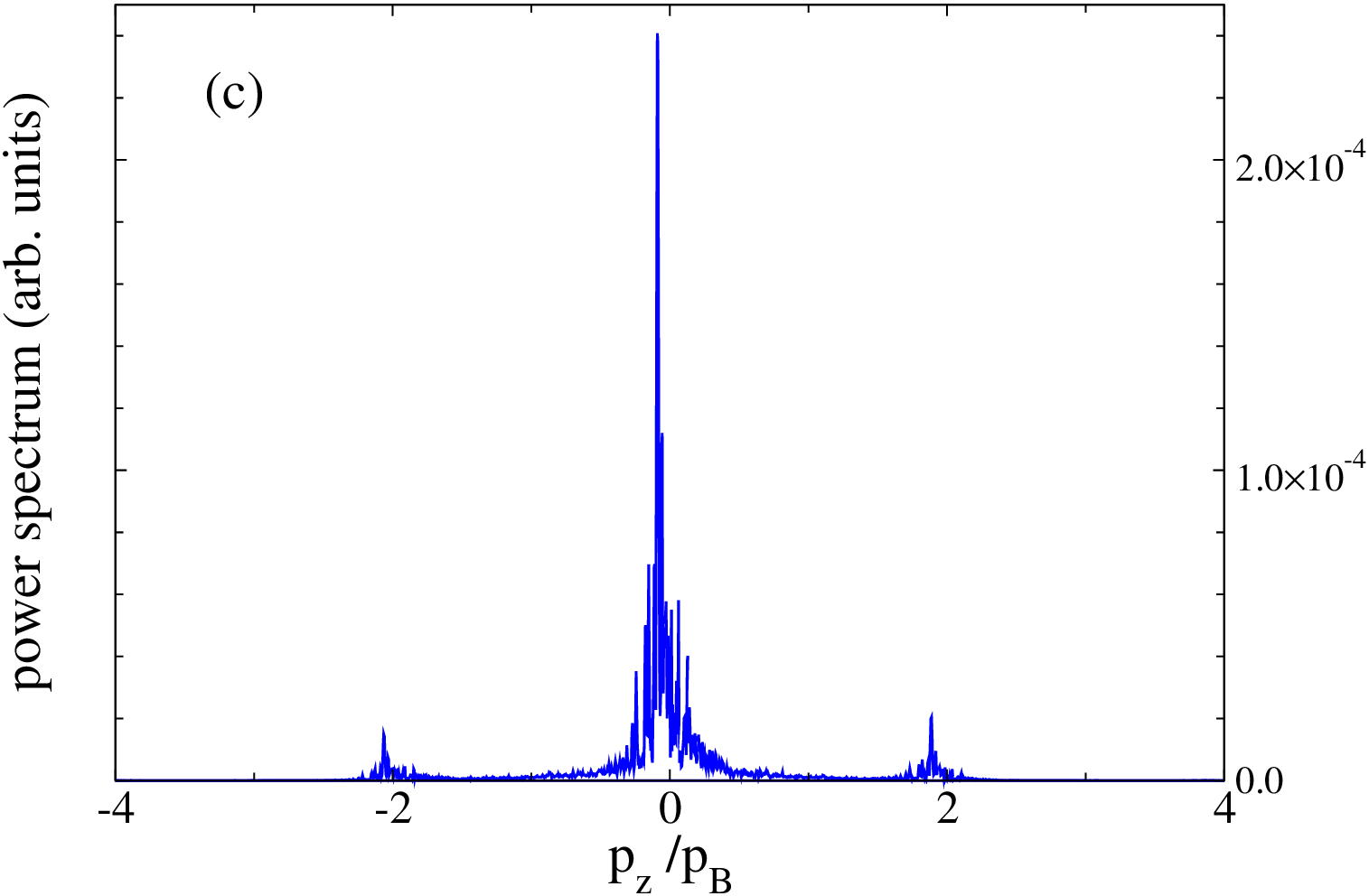}}
\caption{Axial power spectrum of the condensate in the insulator
regime for $t=45$\,ms (a), $60$\,ms (b), and $80$\,ms (c).}
\label{fig:p2}
\end{figure}

Let us now consider the evolution of the axial momentum density
distribution (power spectrum), which provides a deeper understanding
of the system behavior and of the dephasing mechanisms. In
Figs. \ref{fig:p1} and \ref{fig:p2} we show the axial power spectrum
of the condensate, defined as
\begin{equation}
P(p_z)\equiv2\pi\!\int\!\! rdr|\tilde{\Psi}(r,p_z)|^2
\end{equation}
 (the tilde indicates the Fourier transform 
along z), in the case of an initial displacement $\Delta z=40\:\mu$m,
at subsequent evolution times. In particular, in Fig. \ref{fig:p1} we
consider three configurations when the condensate is still in the
superfluid regime, respectively for $t=0$\,ms (a), $15$\,ms (b) and
$30$\,ms (c).  In this case the power spectrum is characterized by
sharp peaks localized at $\tilde{p}_z(t)=\pm2np_B+q(t)$ 
($n=0,\pm1,\dots$, 
here only the zero and first order are visible) \cite{pedri}, 
$p_B=mv_B=\hbar\pi/d$ being the Bragg momentum and
$q(t)$ the condensate quasimomentum.  In fact, when the lattice
intensity is sufficiently high (i.e. we are in the \textit{tight
binding} regime), the condensate in a Bloch state of quasimomentum $q$ can
be written as \cite{ashcroft} (neglecting for simplicity the presence
of the harmonic potential which breaks the periodicity of the lattice
\footnote{This assumption is justified by the fact that the condensate 
extends over many lattice sites, due to the weak trapping along
the axial direction.})
\begin{equation}
\Psi(r,z)=\sum_k{\rm e}^{-iqkd}\varphi(r,z+kd)
\end{equation}
where $\varphi(r,z+kd)$ are wave functions localized at each lattice site
(labeled by the index $k$). It is straightforward to show that in the
coherent regime the momentum distribution is characterized by sharp
peaks whose weight is modulated by the axial Fourier transform of the 
wave function at each lattice site \cite{pedri}
\begin{equation}
P_q(p_z)=
\frac{\sin^2(N_{l}(p_z-q)d/2\hbar)}{\sin^2((p_z-q)d/2\hbar)}
2\pi\!\int\!\! rdr|\tilde{\varphi}(p_z,r)|^2
\end{equation}
$N_l$ being the number of occupied lattice sites \cite{pedri}.

Contrarily, the behavior of the system in the insulator regime is
completely different, as shown in Fig. \ref{fig:p2} for $t=45$\,ms
(a), $60$\,ms (b), and $80$\,ms (c).  Indeed, after the onset of the
dynamical instability, the momentum distribution initially spreads out
losing any signature of coherence (Fig. \ref{fig:p2}(a)). Afterwards,
the system tries to rearrange itself in a coherent fashion as signaled
by the appearance of structures localized in correspondence of the
initial peaks.  Even though these
``peaks'' have a rather large spread (see Fig. \ref{fig:p2}(c)), the
relative population with respect to the central one is of the same
order of that in the full coherent regime (Fig. \ref{fig:p1}(a)).
Notice also that the momentum distribution is centered not exactly in
$p_z=0$ since the condensate is slightly moving toward the trap
center.

\begin{figure}
\centerline{\includegraphics[width=7cm,clip=]{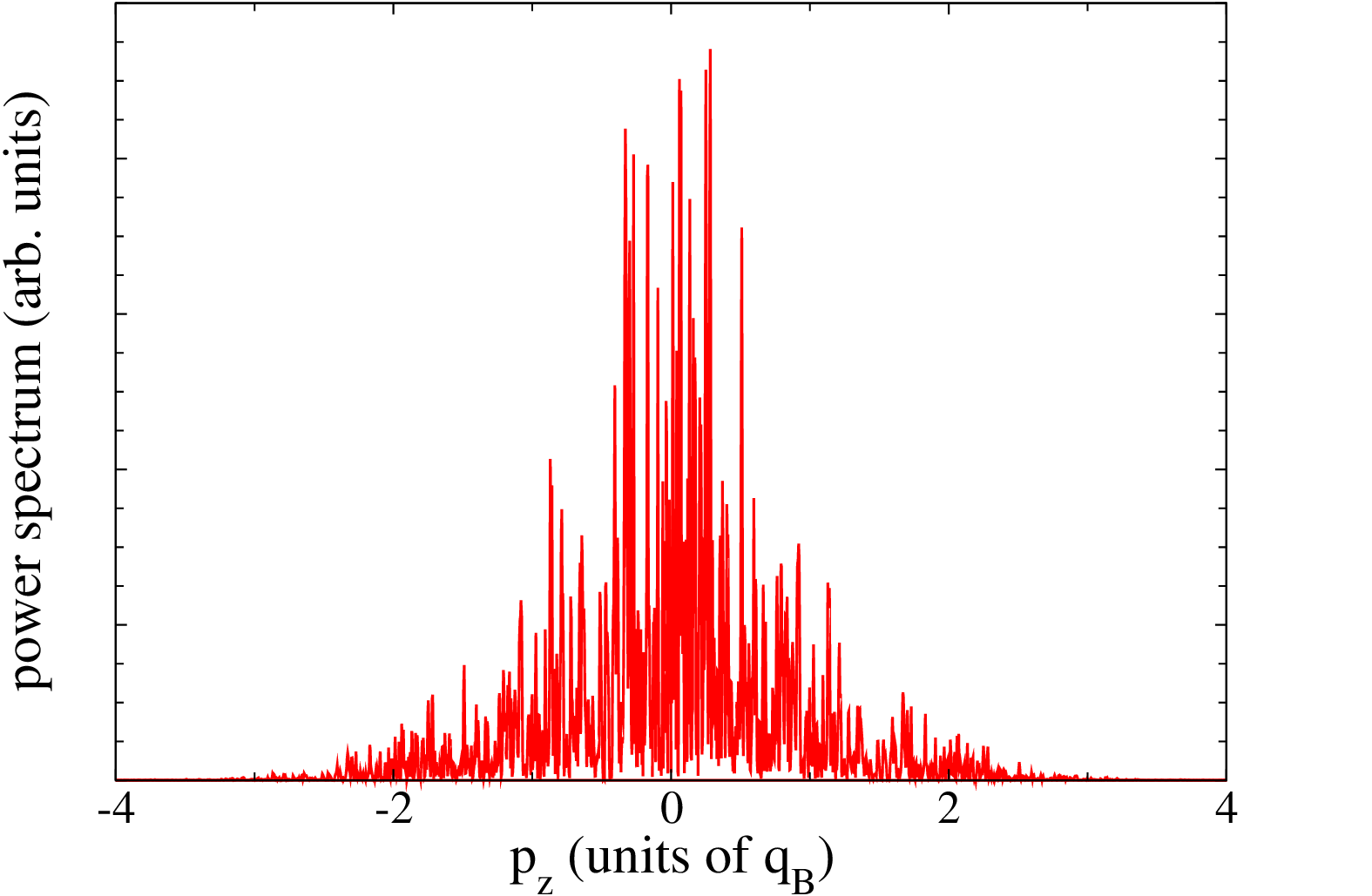}}
\caption{Axial power spectrum of the NPSE solution after an evolution
of $80$\,ms. The distribution is completely spread out indicating a
complete loss of coherence.}
\label{fig:npse}
\end{figure}
For comparison,  in Fig. \ref{fig:npse} we show the axial power
spectrum as obtained from the NPSE, after an evolution of $80$\,ms.
In this case there is no evidence of any localized structure
(contrarily to the case of Fig. \ref{fig:p2}(c)) as further 
confirmation of the fact that
these effective one-dimensional models cannot account for an accurate 
description
of the system behavior in the insulator regime (similar considerations
hold for the GPE-1D).

%-------------------------------------------------
\subsection{Free expansion}
%-------------------------------------------------

In this section we discuss the expected behavior of the system 
after a ballistic expansion. When the system is in the superfluid
regime the internal coherence of the condensate shows up in a clear 
interference pattern characterized by lateral peaks that move
outwards at velocities $v=\pm2p_B/m$ \cite{pedri}.
Contrarily, in the insulator regime 
we expect the overall shape of the interference
pattern and also the expansion of the central peak to be affected by
the partial loss of coherence, as we will show in the following.

To investigate these aspects, instead of solving the full GPE-3D, we
use a simplified model in order to avoid unnecessary heavy numerical
computations \footnote{In principle one should solve the full GPE-3D on a
grid large enough to describe the evolution of the peaks moving
outwards with velocity $\pm2p_B/m$.}. In particular, since we are
mainly interested in the expanded axial profile, we neglect the
contribution of the mean field interaction, and we assume a free
expansion governed by the Schr\"odinger equation
\begin{equation}
i\hbar\frac{\partial}{\partial t} \Psi(r,z;t)=-\frac{\hbar^2\nabla^2}{2m}
\Psi(r,z;t)\,.
\label{eq:schrod}
\end{equation}

The reason to use this approximation is threefold.

(i) First of all, in our case most of the initial energy of the
condensate (at the moment of the release from the trap) is associated
to the fast density modulation, due to the presence of the optical
lattice. Therefore, contrarily to ``usual'' case of a pure harmonic
confinement, now the kinetic energy term predominates over the mean
field one, $E_{mf}\ll E_{kin}$.

(ii) Another reason is that we expect the mean-field interaction to
affect mainly the radial expansion, which is however integrated out in
our treatment.

(iii) Finally, since our aim is to investigate the overall shape of
the interference pattern and to point out the possible presence of
lateral peaks (and/or other signatures of the degree of coherence of
the system) rather than the exact size of the central peak, we can for
this purpose neglect the contribution of the mean field term
\cite{massignan}.

Then, by using Eq. (\ref{eq:schrod}) it is easy to show that the axial density
distribution after an expansion time $t_{exp}$  is given by ($\hbar=1$)
\begin{eqnarray}
\rho(z;t_{exp})&\equiv&
\int\!\! d^2r\left|\Psi(r,z;t_{exp})\right|^2 \\
&=&\int\!\! d^2r\left|\int\!\! dp_z\, e^{i\displaystyle p_z z}
\,\tilde{\Psi}_0(r,p_z)\,e^{-i(p_z^2/2m) t_{exp}}\right|^2\!\!,
\nonumber
\end{eqnarray}
$\tilde{\Psi}_0(r,p_z)$ being the axial Fourier transform of the
initial wave function (at the time of the release from the trap).  
\setlength{\parskip}{0.6ex minus 0.3ex}

\begin{figure}
\centerline{\includegraphics[width=8cm,clip=]{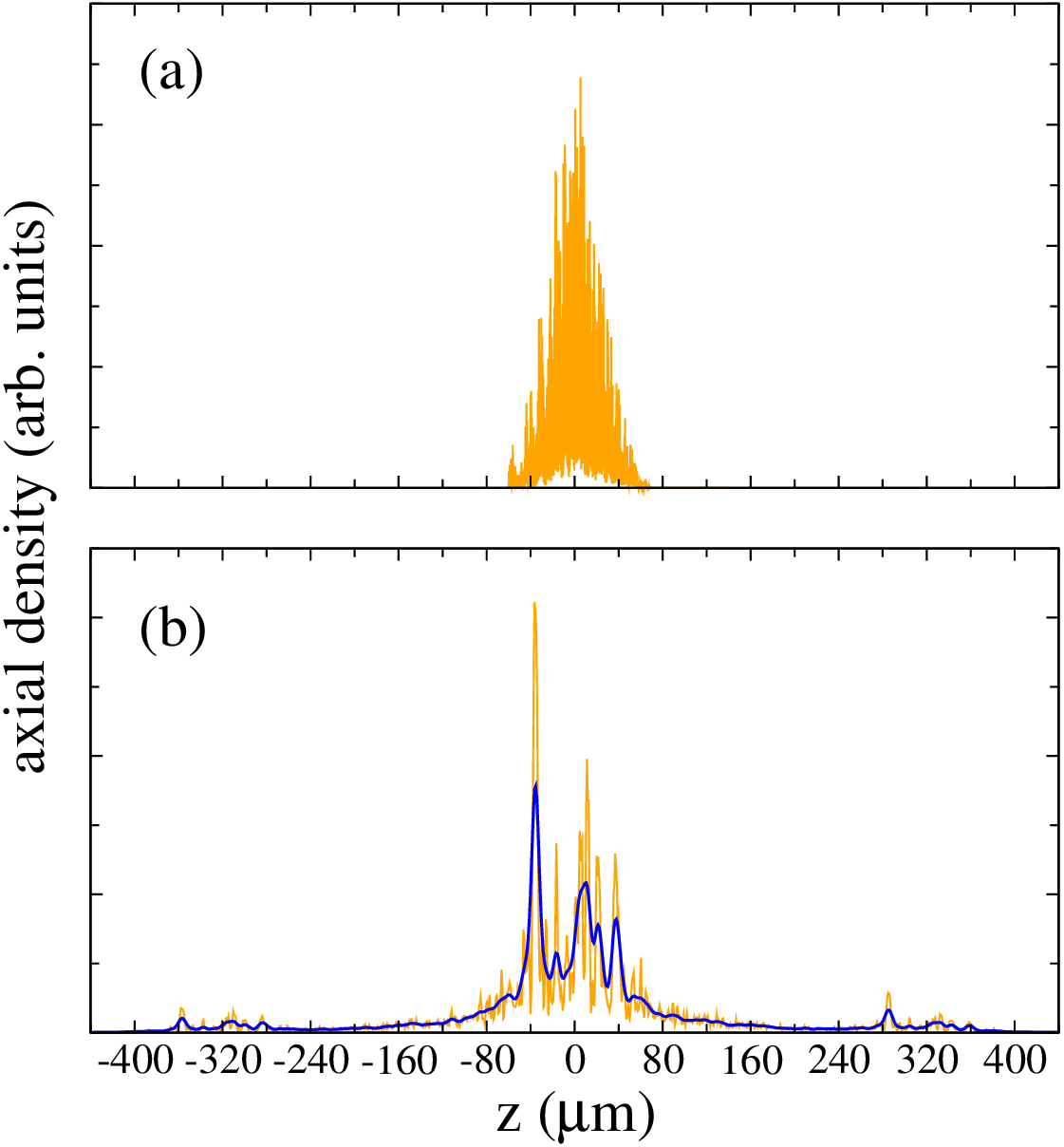}}
\caption{
(a) Axial profile of the condensate in the trap, in the
insulator regime (for a displacement $\Delta z=40\,\mu$m and after
70\,ms of trapped evolution).  (b) Interference pattern (lateral
peaks) and fringes (in the central peak) after a subsequent expansion
of about 30\,ms. The marked line in (b) is obtained by taking into
account the finite resolution of the imaging apparatus as described in
the text.}
\label{fig:exp}
\end{figure}

As an example in Fig. \ref{fig:exp} we show the typical shape of the
density distribution before (a) and after (b) an expansion of about
30\,ms, corresponding to a case similar to those in
Figs. \ref{fig:p2}(b)-(c). We see that both distributions are
characterized by a rather fragmented profile, although after the
expansion (Fig. \ref{fig:exp}(b)) it is possible to identify some
 structures localized in correspondence of the first order peaks of the
analogous expanded profile in the case of a fully coherent condensate
(at $z\simeq320\,\mu$m) \cite{pedri}.

Furthermore, to take into account the effect of the finite resolution
of the experimental imaging apparatus, in Fig. \ref{fig:exp}(b) we
also show the convolution (marked line) of the axial density with a
gaussian distribution of width $\sigma$ (here $2\sigma=6\,\mu$m,
according to the typical experimental resolution). This procedure
clearly evidences the persistence of lateral peaks and the appearance
of fringes in the central one, whose presence is a signature of the
partial loss of coherence (as the reduced visibility of the lateral
peaks).

Note that although the actual shape of the interference pattern
depends on the initial conditions (as for example the trapped
evolution time), the overall behavior shown in Fig. \ref{fig:exp}(b)
can be reproduced for a wide range of trap displacements and
evolution/expansion times.  We have also verified that the main
features discussed in this work hold for lattice intensities in the
range $2<s<10$.

Both the persistence of the lateral peaks and the appearance of 
fringes in the central peak in the density distribution of the 
expanded condensate have been recently observed in the experiments at 
LENS \cite{lens,fort}.

%===============================================================
\section{Conclusions}
\label{sec:conclusions}
%===============================================================

We have studied the phase coherence of a Bose-Einstein condensate that
undergoes a dynamical superfluid-insulator transition during the
trapped evolution in the presence of a one-dimensional optical
lattice, as recently observed at LENS \cite{lens,fort} and discussed
in \cite{smerzi,wuniu,adhikari}.

From the comparison of the solution of the three-dimensional
Gross-Pitaevskii equation with that of two effective 1D models, we
have demonstrated that the inclusion of the transverse degrees of
freedom is crucial to account for the actual behavior of the system,
as observed in the experiments \cite{lens,fort}. In particular we have
shown that the breakdown of the superfluid current is not associated
to a sharp transition as predicted for the pure one-dimensional case,
but there exists a range of parameters for which the condensate can
partially recover some coherence during the subsequent evolution.
 
We have also shown that the degree of coherence of the system affects
significantly the interference pattern after the ballistic expansion
of the condensate, characterized by the persistence of lateral
peaks (as a signature of a partial coherence) and by the appearance 
of fringes in the central peak (due to the dephasing of the system).

These results open interesting questions about the precise role played
by the radial degrees of freedom on the excitation spectrum and on the 
decoherence mechanism. 
The investigation of these aspects requires the analysis
of the 3D Bogoliubov spectrum of the system, and will be addressed in a
future publication.

%===============================================================
\begin{acknowledgments}
We acknowledge stimulating discussions with F. Dalfovo,
L. Pitaevskii and C. Fort.
\end{acknowledgments}
%===============================================================


\begin{thebibliography}{}
\bibitem{sites} 
For the numerical grid we use up to 64 sites in the radial
direction and 8192 in the axial one (the direction of the lattice).
\end{thebibliography}
\end{document}